%% file: draft.tex
\documentclass[prd,twocolumn,tightenlines,showpacs,amsmath,amssymb]{revtex4-1}
\usepackage{amssymb}
\usepackage{multirow}
\usepackage{graphicx,epsfig}
\usepackage{dcolumn}
\usepackage{bm}
\usepackage[dvipdfm,CJKbookmarks=true,unicode,colorlinks,linkcolor=blue,anchorcolor=blue,citecolor=blue,pdfborder={0 0 0}]{hyperref}
\usepackage{amsfonts}
\usepackage{keyval,graphicx}
\usepackage{textcomp,wasysym}
\usepackage{xspace}

\begin{document}
\newcommand{\cinst}[2]{$^{\mathrm{#1}}$~#2\par}
\newcommand{\crefi}[1]{$^{\mathrm{#1}}$}
\newcommand{\crefii}[2]{$^{\mathrm{#1,#2}}$}
\newcommand{\crefiii}[3]{$^{\mathrm{#1,#2,#3}}$}
\newcommand{\HRule}{\rule{0.5\linewidth}{0.5mm}}
\newcommand{\br}[1]{\mathcal{B}#1}
\newcommand{\el}[1]{\mathcal{L}#1}
\newcommand{\ef}[1]{\mathcal{F}#1}
\newcommand{\etapr}{\ensuremath{\eta^\prime}\xspace}



\title{\boldmath  Observation of $\eta^{\prime}\to\omega e^{+} e^{-}$}

\input{authors_jul2015}

\begin{abstract}
  Based on a sample of \etapr mesons produced in the radiative decay
  $J/\psi\to\gamma\eta^{\prime}$ in $1.31\times 10^9$
  $J/\psi$ events collected with the BESIII detector, the decay
  $\eta^{\prime}\to\omega e^{+} e^{-}$ is observed for the first time,
  with a statistical significance of $8\sigma$.
  The branching fraction is
  measured to be $\mathcal{B}(\eta^{\prime}\to\omega e^{+} e^{-})=
  (1.97\pm0.34(\text{stat})\pm0.17(\text{syst}))\times10^{-4}$, which is in
  agreement with theoretical predictions.  The branching fraction of
  $\eta^{\prime}\to\omega\gamma$ is also measured to be
  $(2.55\pm0.03(\text{stat})\pm0.16(\text{syst}))\times10^{-2}$, which is the
  most precise measurement to date, and the relative branching
  fraction $\frac{\mathcal{B}(\eta^{\prime}\to \omega e^{+}
    e^{-})}{\mathcal{B}(\eta^{\prime}\to \omega \gamma)}$ is
  determined to be
  $(7.71\pm1.34(\text{stat})\pm0.54(\text{syst}))\times10^{-3}$.

\end{abstract}



\maketitle

\section{Introduction}
The main decays of the ${\eta}^{\prime}$ meson~\cite{PDG} fall into
two distinct classes.  The first class consists of hadronic decays into three pseudoscalar mesons, such as
${\eta}^{\prime}$ ${\to}$ ${\eta}{\pi}{\pi}$, while the second class
has radiative decays into vector particles with quantum number
$J^{PC} = 1^{--}$, such as ${\eta}^{\prime} \to \gamma \gamma,
\rho \gamma$, or $\omega \gamma $. Model-dependent approaches for
describing low energy mesonic interactions, such as vector meson
dominance (VMD)~\cite{Phys.Rev.C61.i}, and the applicability of
chiral perturbation theory~\cite{Phys.Rev.C61.i} can be tested in
${\eta}^{\prime}$ decays.

It is of interest to study the decay ${\eta}^{\prime}$ ${\to}$ $Ve^{+}e^{-}$ (V
represents vector meson) which proceeds via a two-body radiative decay
into a vector meson and an off-shell photon. The electron-positron
invariant mass distribution provides information about the intrinsic
structure of the ${\eta}^{\prime}$ meson and the momentum dependence
of the transition form factor. Recently, BESIII reported the
measurement of $\eta^\prime\to \pi^+\pi^-
e^{+}e^{-}$~\cite{Phys.Rev.D87.092011}, which is found to be
dominated by ${\eta}^{\prime}$ ${\to}\rho e^{+}e^{-}$, in agreement
with theoretical predictions~\cite{Phys.Rev.C61.i,Borasoy:2007pr}.

Based on theoretical models~\cite{Phys.Rev.C61.i,Eur.Phys.J.A48.190},
the branching fraction of ${\eta}^{\prime}$ ${\to}$
${\omega}e^{+}e^{-}$ is predicted to be around $2.0{\times}10^{-4}$,
but until now there has been no measurement of this decay. A sample of
$1.31\times 10^9$ $J/\psi$ events ($2.25\times10^8$
events~\cite{njpsi} in 2009 and $1.09\times10^9$~\cite{njpsi2} in
2012) has been collected with the BESIII detector and offers us a
unique opportunity to investigate $\eta^\prime$ decays via
$J/\psi\to\gamma\eta^\prime$. In this paper, the observation of
$\eta^\prime \rightarrow \omega e^+e^-$, the analysis of the
decay $\eta^{\prime}\to\omega\gamma$, and the ratio of their branching
fractions are reported.

\section{Detector and Monte Carlo simulation}

The BESIII detector is a magnetic spectrometer located
at the Beijing Electron Positron Collider (BEPCII,~\cite{BEPCII}), which is a
double-ring $e^+ e^-$ collider with a design peak luminosity of
$10^{33}$ cm$^{-2}$ s$^{-1}$ at a center-of-mass energy of 3.773
GeV. The cylindrical core of the BESIII detector consists of a
helium-based multilayer drift chamber (MDC), a plastic scintillator
time-of-flight system (TOF), and a CsI (Tl) electromagnetic
calorimeter (EMC), which are all enclosed in a superconducting
solenoidal magnet providing a 1.0 T (0.9 T for the 2012 run period)
magnetic field. The solenoid is supported by an octagonal flux-return
yoke with modules of resistive plate muon counters (MUC) interleaved
with steel. The acceptance for charged particles and photons is 93\%
of the full 4$\pi$ solid angle. The momentum resolution for charged
particles at 1 GeV/$c$ is 0.5\%, and the resolution of the ionization
energy loss per unit path-length ($dE/dx$) is 6\%. The EMC measures
photon energies with a resolution of 2.5\% (5\%) at 1 GeV in the
barrel (end-caps). The time resolution for the TOF is 80 ps in the
barrel and 110 ps in the end-caps. Information from
  the TOF and $dE/dx$ is combined to perform particle identification
  (PID).

The estimation of backgrounds and the determinations of detection
efficiencies are performed through Monte Carlo (MC) simulations. The
BESIII detector is modeled with \textsc{geant}{\footnotesize
  4}~\cite{Agostinelli:2003hh,Allison:2006ve}. The production of the
$J/\psi$ resonance is implemented with the MC event generator
\textsc{kkmc}~\cite{Jadach:1999vf, Jadach:2000ir}, while the decays
are simulated with \textsc{evtgen}~\cite{EvtGen}. Possible backgrounds
are studied using a sample of `inclusive' $J/\psi$ events
of approximately the equivalent luminosity of data, in which the
known decays of the $J/\psi$ are modeled with branching fractions
being set to the world average values from the Particle Data Group
(PDG)~\cite{PDG}, while the remaining decays are generated with the
\textsc{lundcharm} model~\cite{Chen:2000}.
For this analysis, a signal MC sample ($6.0\times10^5$ events), based on the VMD model and
chiral perturbation theory~\cite{Phys.Rev.C61.i} for $J/\psi\to\gamma\eta^{\prime}$,
$\eta^{\prime}\to\omega e^{+} e^{-}$, $\omega\to\pi^{0}\pi^+\pi^-$,
$\pi^{0}\to\gamma\gamma$, is generated to optimize the selection
criteria and determine the detection efficiency.

\section{\boldmath ANALYSIS of $J/\psi\to\gamma\eta^\prime$ }

In this analysis, the $\etapr$ meson is produced in the radiative decay
$J/\psi\to\gamma\eta^\prime$. The $\omega$ meson is observed in its dominant
$\pi^+\pi^-\pi^0$ decay mode, and the $\pi^0$ is detected in
$\pi^0\to\gamma\gamma$. Therefore, signal events are observed in the
topology $\gamma\gamma\gamma\gamma\pi^+\pi^-$ for the
$\eta^{\prime}\to\omega\gamma$ mode, and
$\gamma\gamma\gamma\pi^+\pi^-e^+e^-$ for $\eta^{\prime}\to\omega
e^{+}e^{-}$. We apply the following basic reconstruction and selection
criteria to both channels:

We select tracks in the MDC within the polar angle range
$|\cos\theta|<0.93$ and require that the points of closest approach to
the beam line be within $\pm 20$~cm of the interaction point in the
beam direction and within $2$~cm in the plane perpendicular to the
beam.

Photon candidates are reconstructed by clustering signals in EMC
crystals. At least four photon candidates are required, and the
minimum energy of each must be at least $25$~MeV for barrel showers
($|\cos\theta| < 0.80$) and $50$~MeV for endcap showers
($0.86<|\cos\theta|<0.92$).  To exclude showers due to the
bremsstrahlung of charged particles, the angle between the nearest
charged track and the shower must be greater than $10^{\circ}$.
To suppress electronics noise and energy deposits unrelated to the event,
the EMC cluster time is restricted to be within a 700 ns window near the event start time.

\color{black}

\subsection{\boldmath $\eta^{\prime}\to\omega\gamma$}\label{etaptogammaomega}
For the decay $\eta^{\prime}\to\omega\gamma$, two particles with opposite
charge are required. No particle identification (PID) is used, and the
two tracks are taken to be positive and negative pions from the $\omega$.

\begin{figure}[htbp]
\includegraphics[width=0.45\textwidth]{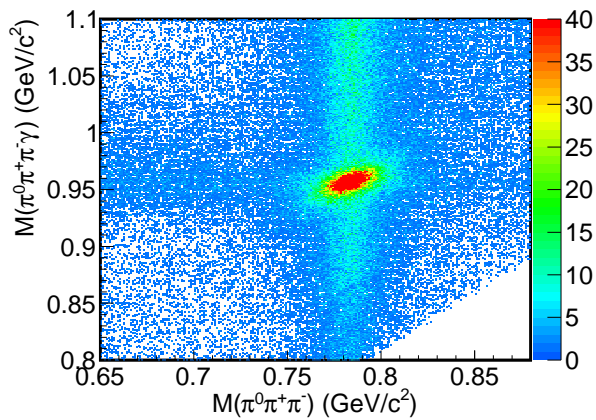}

\caption{Distribution of $M(\pi^{0}\pi^{+}\pi^{-}\gamma)$ versus
  $M(\pi^{0}\pi^{+}\pi^{-})$ from data. \label{fit1}}
\end{figure}

A four-constraint (4C) kinematic fit imposing energy-momentum
conservation is performed under the hypothesis of
$J/\psi\to\gamma\gamma\gamma\gamma\pi^{+}\pi^{-}$.  If there are more
than four photons, the combination with the smallest
$\chi^{2}_{\gamma\gamma\gamma\gamma\pi^{+}\pi^{-}}$ is retained.
Events with $\chi^{2}_{\gamma\gamma\gamma\gamma\pi^{+}\pi^{-}} < 80$
are retained for further analysis. Since $J/\psi\rightarrow\gamma\eta^\prime$ is a two-body decay, the
radiative photon carries a unique energy of 1.4 GeV. Hence the photon
with maximum energy is taken as the radiative photon,
and its energy is required to be greater than 1.0 GeV.
The photon pair combination with $\gamma\gamma$
invariant mass closest to the $\pi^{0}$ mass is considered as the
$\pi^0$ candidate in the final state, and its invariant mass must
satisfy $|M(\gamma\gamma)-M_{\pi^{0}}|<0.015$
GeV$/c^{2}$, where $M_{\pi^{0}}$ is the world average value of the
$\pi^{0}$ mass~\cite{PDG}. With these requirements, the decay
$\eta^\prime\to\omega\gamma$ is observed in the distribution
of $M(\pi^{0} \pi^{+} \pi^{-} \gamma )$ versus $M( \pi^{0} \pi^{+}
\pi^{-})$, shown in Fig.~\ref{fit1}. Besides the region of interest in
Fig.~\ref{fit1} , there is a vertical band around the $\omega$ mass
region, which comes from $J/\psi \to \omega \eta$, $\omega
\pi^{0}$ and $\omega \pi^0\pi^0$ background, while a horizontal band
also exists around the $\eta^{\prime}$ mass region, which comes from
$J/\psi\to\gamma\eta^{\prime}$, $\eta^{\prime}\to\eta\pi^{+}\pi^{-}$
and $\gamma\rho^{0}$.

\begin{figure}[htbp]
    \includegraphics[width=0.45\textwidth]{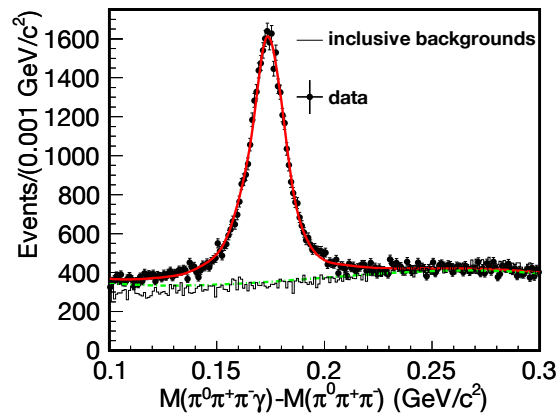}
    \caption{Distribution of the mass difference
      $M(\pi^0\pi^+\pi^-\gamma)-M(\pi^0\pi^+\pi^-)$. The dots with
      error bars are data, the histogram shows the MC simulation of inclusive $J/\psi$
      decays. The solid curve represents the fit results, and the
      dashed curve is the background determined by the fit.\label{fit2}}
\end{figure}

To improve the mass resolution, as well as to better handle the
background in the vertical band around the $\omega$ mass region and
horizontal band around the $\eta^{\prime}$ mass region, we determine
the signal yield from the distribution of the difference between
$M(\pi^{0} \pi^{+} \pi^{-} \gamma )$ and $M( \pi^{0} \pi^{+}
\pi^{-})$. The backgrounds in the vertical and horizontal bands do not
peak in the signal region, which is demonstrated by the inclusive MC
sample, as shown by the histogram in Fig.~\ref{fit2}.

To determine the signal yield, an unbinned maximum likelihood fit to
the mass difference $M(\pi^{0} \pi^{+} \pi^{-} \gamma ) - M( \pi^{0}
\pi^{+} \pi^{-})$ is performed, in which the signal shape is described
by the MC shape convoluted with a Gaussian function to account for the
difference in resolution between data and MC simulation, and the
background is described by a 3rd-order Chebychev
polynomial. $33187\pm351$ $\eta^\prime\to\omega\gamma$ signal events
are obtained from the fit, whose curve is shown in
Fig.~\ref{fit2}. With the detection efficiency, $(21.87\pm0.02)\%$,
obtained from MC simulation, the branching fraction,
$(2.55\pm0.03)\times10^{-2}$, listed in Table~\ref{sum}, is
determined.

\subsection{\boldmath  $\eta^{\prime}\to\omega e^{+}e^{-}$}

For $\eta^{\prime}\to\omega e^{+}e^{-}$ decay, candidate events with
four well-reconstructed charged tracks and at least three photons are
selected. The charged track and good photon selections are exactly the
same as described above.

To select candidate events and select the best photon combination when
additional photons are found in an event, the combination with the
smallest $\chi^2_{4C + \text{PID}}$ is retained. Here $\chi^{2}_{4C +
\text{PID}}=\chi^{2}_{\rm 4C} +
\sum_{j=1}^{4}\chi^{2}_{\rm PID}(j)$ is the sum of the chi-squares
from the $\rm 4C$ kinematic fit and from PID, formed
by combining TOF and $dE/dx$ information of each charged track for
each particle hypothesis (pion, electron, or muon).  If the combination
with the smallest $\chi^{2}_{4C+\text{PID}}$ corresponds to two oppositely
charged pions and an electron and positron, and has $\chi^{2}_{\rm
4C}< 80$, the event is kept as a
$\gamma\gamma\gamma\pi^{+}\pi^{-}e^{+}e^{-}$ candidate.  As
in the analysis of $\eta^{\prime}\to\omega\gamma$, the selected photon
with maximum energy is taken as the radiative photon, and its energy
is required to be greater than 1.0~GeV. The other two photons are
further required to be consistent with a $\pi^0$ candidate,
$|M(\gamma\gamma)-M_{\pi^{0}}|<0.015$ GeV$/c^{2}$.

\begin{figure*}[htbp]
    \includegraphics[width=0.3\textwidth]{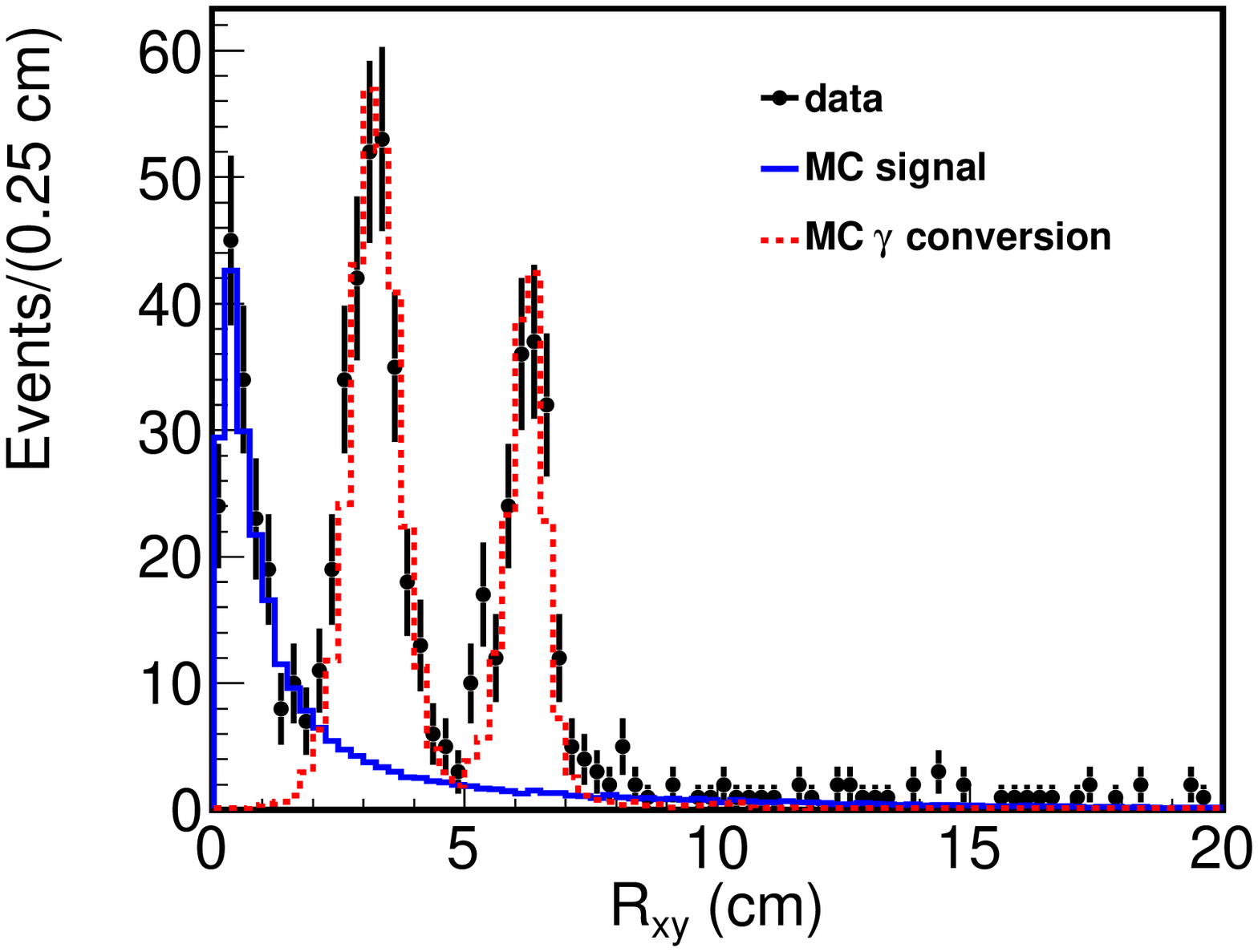}
    \put(-35,95){(a)}
    \includegraphics[width=0.3\textwidth]{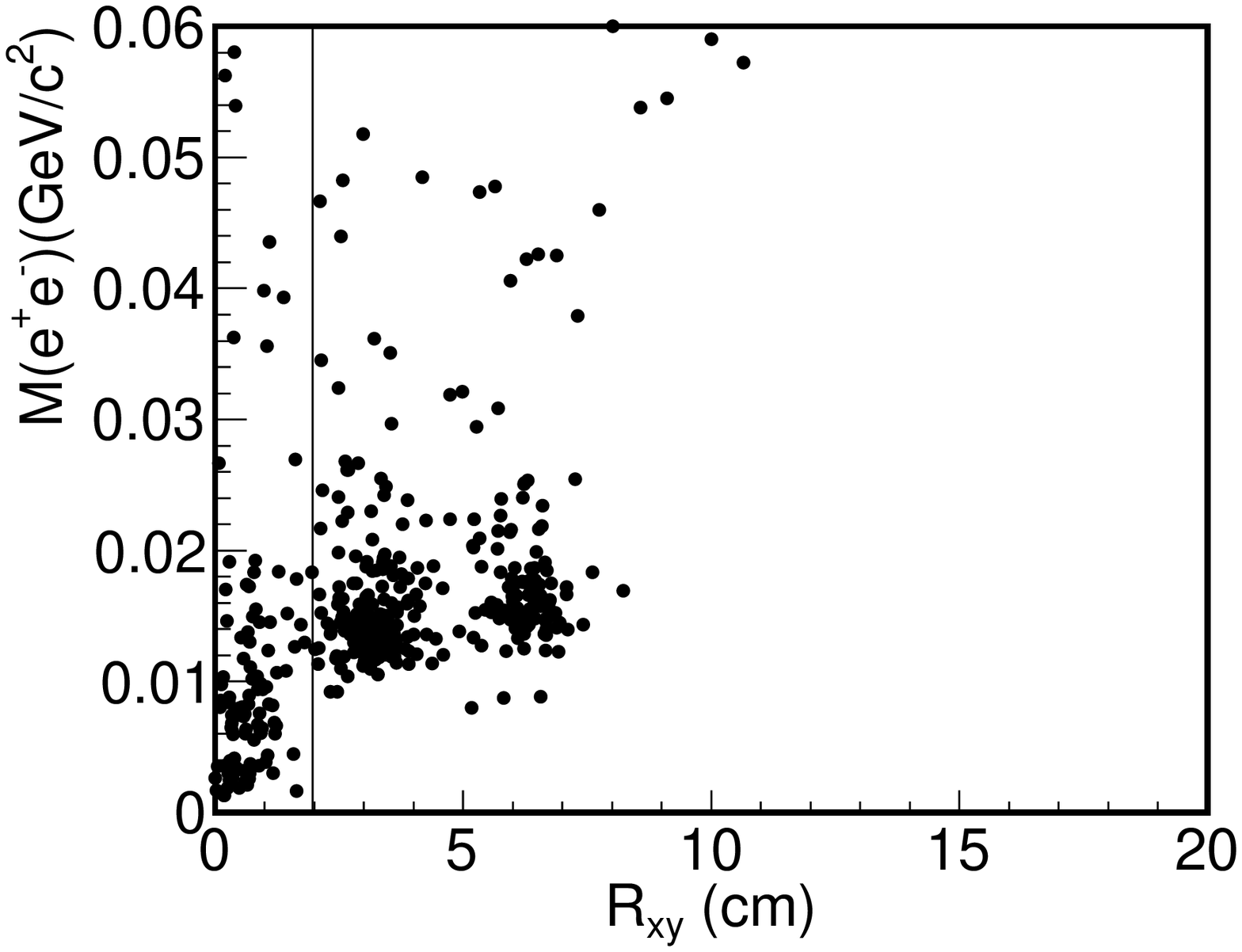}
    \put(-35,95){(b)}
    \includegraphics[width=0.3\textwidth]{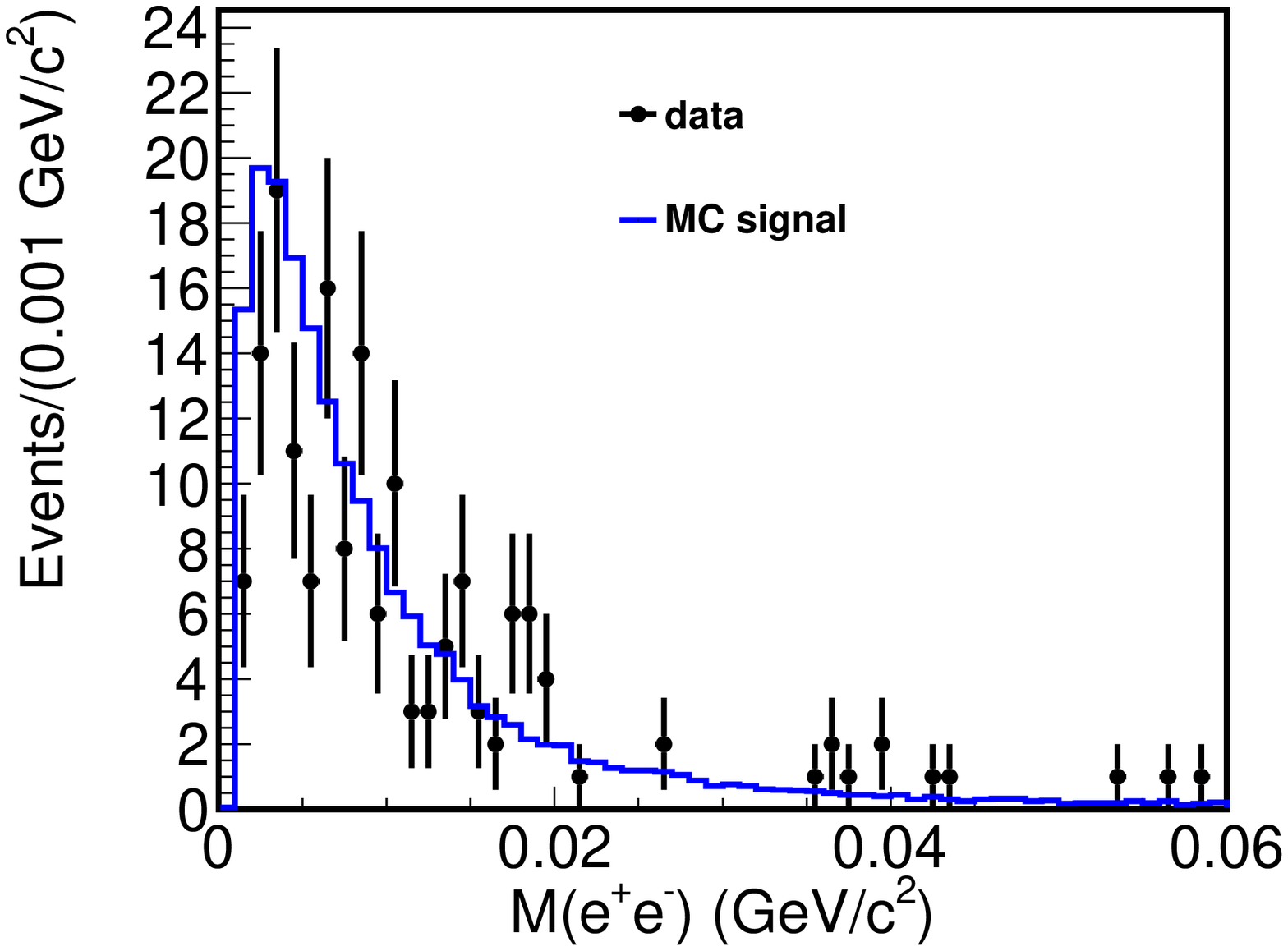}
    \put(-35,95){(c)}
    \caption{(a)~Distribution of the distance of the reconstructed
      $e^+e^-$ vertex from the $z$ axis, $R_{xy}$, where the dots with
      error bars are data, the solid histogram is signal MC
      simulation, and the dotted histogram is MC simulation of
      $\eta^{\prime}\to\omega\gamma$.  (b)~Distribution of
      $M(e^{+}e^{-})$ versus $R_{xy}$, where the requirement of $R_{xy}<2
      $ cm is indicated as the vertical line.  (c)~Distribution of
      $M(e^{+}e^{-})$ with the requirement $R_{xy}<2$ cm, where the dots with error
      bars are data and the solid histogram is signal MC
      simulation.}\label{rxy}
\end{figure*}

With the above selection criteria, MC simulation shows that
background peaking under the signal comes from $J/\psi\to\gamma\eta^{\prime}$,
$\eta^{\prime}\to\omega\gamma$, with the $\gamma$ from the $\eta^{\prime}$
decay subsequently converting to an electron-positron pair. The
distribution of the distance from the reconstructed vertex point of an
electron-positron pair to the $z$ axis, defined as $R_{xy}$, is shown
in Fig.~\ref{rxy}~(a). As expected from MC simulation of
$J/\psi\to\gamma\eta^{\prime}$, $\eta^{\prime}\to \omega\gamma$, the
peaks around $R_{xy}=3$~cm and $R_{xy}=6$~cm match the position of the
beam pipe and the inner wall of the MDC, respectively, as shown in
Fig.~\ref{rxy}~(a). From the distribution of $M(e^{+}e^{-})$ versus
$R_{xy}$ and the $M(e^{+}e^{-})$ projections, shown in
Figs.~\ref{rxy}~(b) and (c), the requirement of $R_{xy}<2$ cm can
cleanly discriminate signal from the background. The number of peaking
background events from $\eta^\prime\to\omega\gamma$ that still survive
is estimated to be $2.6\pm0.3$ from MC simulation taking the
branching fraction of $J/\psi \to\gamma \eta^{\prime}, \eta^{\prime}
\to \omega \gamma$ from this analysis, where the error is
statistical. This background will be subtracted in the calculation of
the branching fraction of $\eta^\prime\to \omega e^+e^-$.

\begin{figure*}[htbp]
   \includegraphics[width=6cm]{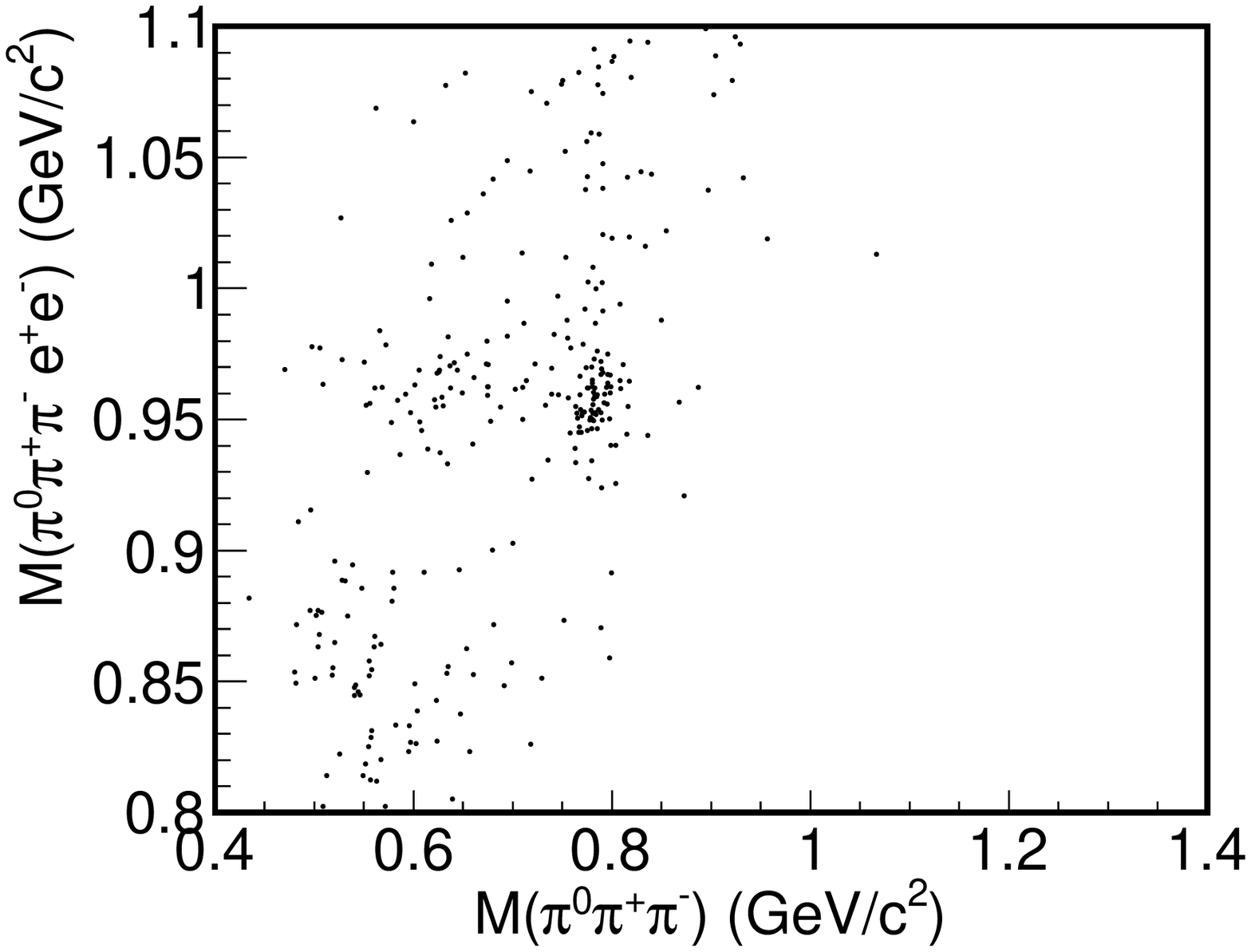}
   \put(-35,105){(a)}
   \includegraphics[width=6cm]{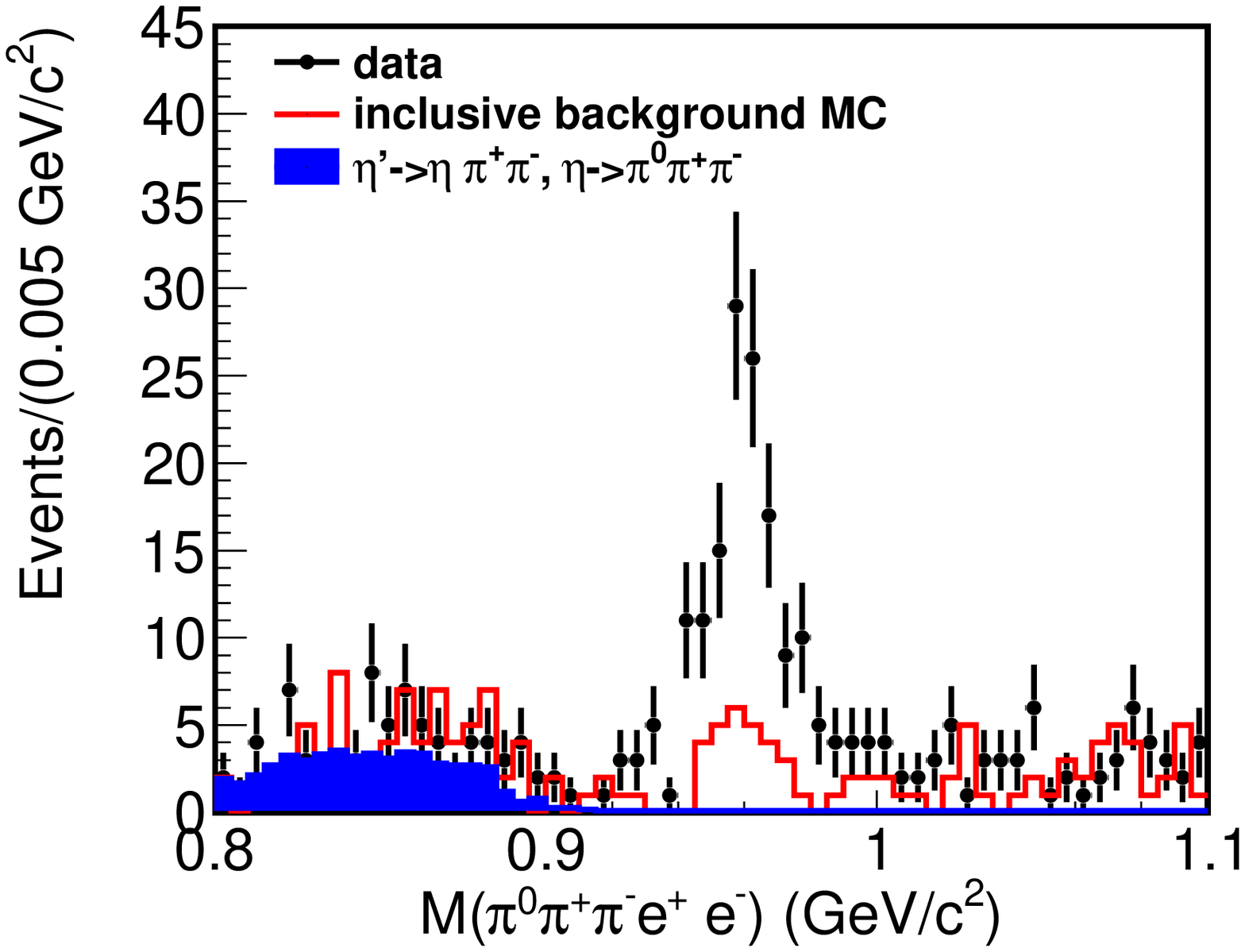}
   \put(-35,105){(b)}

   \includegraphics[width=6cm]{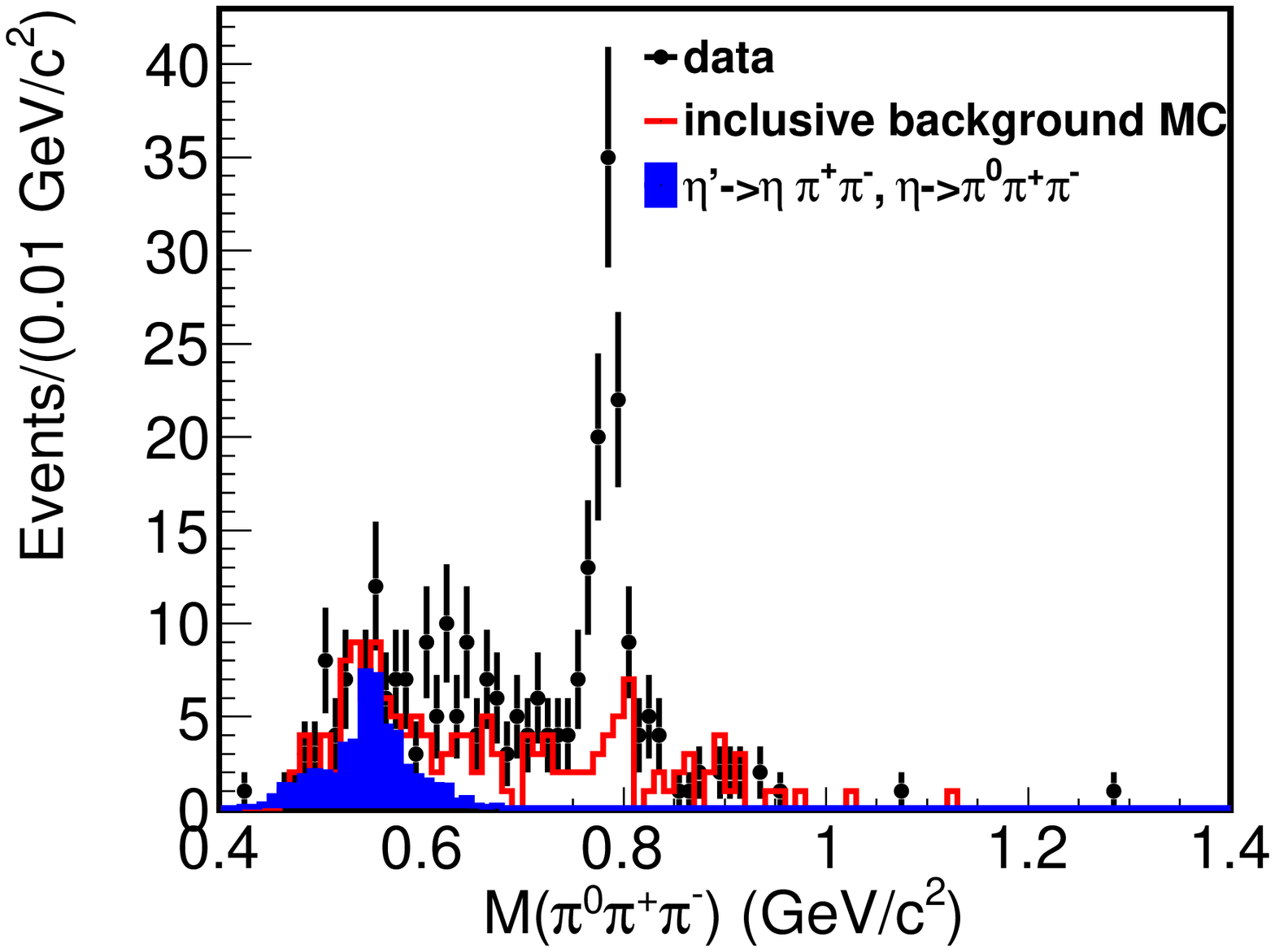} \put(-35,106){(c)}
   \includegraphics[width=6cm]{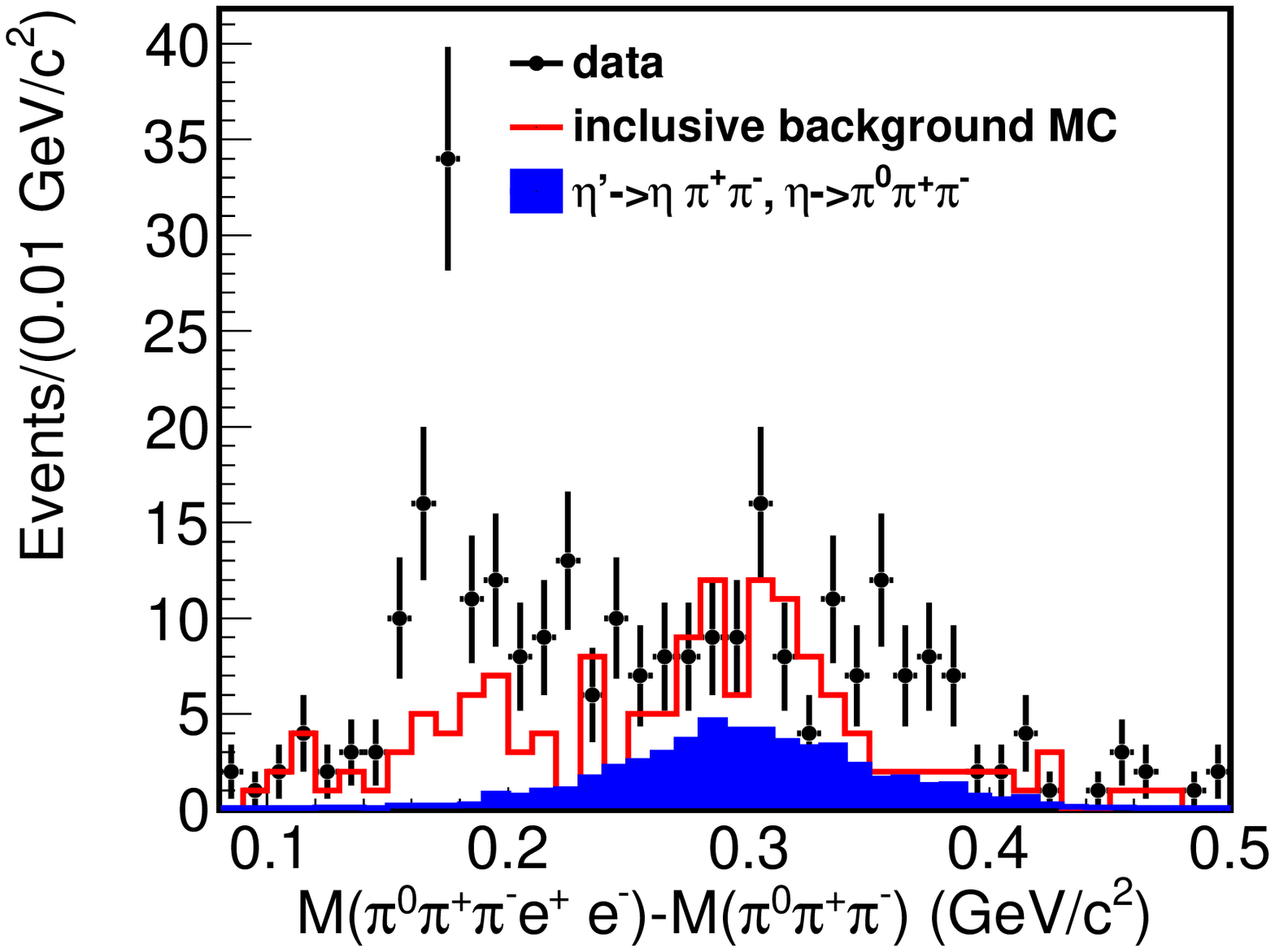}
   \put(-35,105){(d)} \caption{ (a)~Distribution of
   $M(\pi^0\pi^+\pi^-e^+e^-)$ versus $M(\pi^0\pi^+\pi^-)$. (b)~Invariant
   mass spectrum of $\pi^0\pi^+\pi^-e^+e^-$. (c)~Invariant mass spectrum
   of $\pi^0\pi^+\pi^-$. (d)~Distribution of $M(\pi^0\pi^+\pi^-
   e^{+}e^{-})-M(\pi^0\pi^+\pi^-)$. The solid
   histogram represents the remaining events from the inclusive MC
   sample, and the shaded histogram shows misidentified events from the
   background channel $\eta^{\prime}\to\eta\pi^{+}\pi^{-}$ normalized
   by using the branching fractions from the PDG~\cite{PDG}.
\label{etapvsomega}}
\end{figure*}

With all the above selection criteria being applied,
the scatter plot of $M(\pi^{0}\pi^{+}\pi^{-}e^{+}e^{-})$ versus
$M(\pi^{0}\pi^{+}\pi^{-})$ is shown in Fig.~\ref{etapvsomega}~(a),
where the cluster in the $\eta^{\prime}$ and $\omega$ region
corresponds to the decay $\eta^{\prime}\to \omega e^{+}e^{-}$. The
$\eta^\prime$ and $\omega$ peaks are clearly seen in the distributions
of $M(\pi^{0}\pi^{+}\pi^{-}e^{+}e^{-})$ (Fig.~\ref{etapvsomega}~(b))
and $M(\pi^{0}\pi^{+}\pi^{-})$ (Fig.~\ref{etapvsomega}~(c)),
respectively.

The same selection is applied to the inclusive MC sample of $1.2\times 10^9$ $J/\psi$ events
to investigate possible background channels. The corresponding normalized
distributions of $M(\pi^{0}\pi^{+}\pi^{-}e^{+}e^{-})$ and
$M(\pi^{0}\pi^{+}\pi^{-})$ are shown as the histograms in
Fig.~\ref{etapvsomega}~(b) and (c).  One of the dominant backgrounds
is from events with multiple $\pi^0$ in the final state with one
$\pi^0$ undergoing Dalitz decay to $\gamma e^+e^-$. Another important
background, $\eta^\prime\to\pi^+\pi^-\eta$, $\eta\to\pi^0\pi^+\pi^-$
with the pion pair from the $\eta'$ decay misidentified as an
electron-positron pair, produces an accumulation at the low mass
region in the distributions of $M(\pi^{0}\pi^{+}\pi^{-}e^{+}e^{-})$
and $M(\pi^{0}\pi^{+}\pi^{-})$, and at the high mass region in
$M(\pi^{0}\pi^{+}\pi^{-}e^{+}e^{-})$-$M(\pi^{0}\pi^{+}\pi^{-})$, which
is shown as the shaded histograms in Fig.~\ref{etapvsomega}~(b), (c)
and (d), normalized with the branching fraction from the PDG.  The
distribution of $M(\pi^0\pi^+\pi^- e^{+}e^{-})$ - $M(\pi^0\pi^+\pi^-)$
is shown in Fig.~\ref{etapvsomega}~(d).  From this study of the
inclusive MC sample, no peaking background events are expected.

\begin{figure}[htbp]
\includegraphics[width=0.45\textwidth]{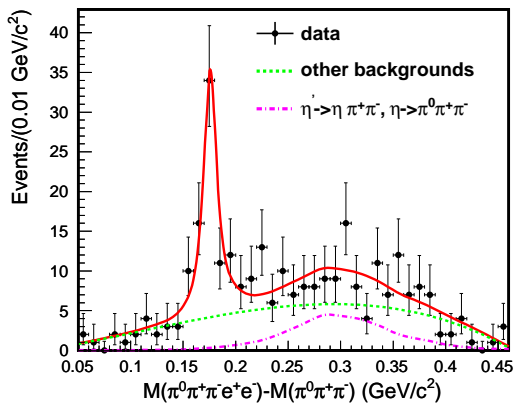}
\caption{ Distribution of $M(\pi^0\pi^+\pi^- e^{+}
  e^{-})-M(\pi^0\pi^+\pi^-)$ and the fit results. The crosses show the
   distribution of data. The dash-dotted line represents the
   $\eta^\prime\to\pi^+\pi^-\eta$ component, and the dotted curve
   shows the background except
   $\eta^\prime\to\pi^+\pi^-\eta$.\label{fit3}}
\end{figure}

To determine the $\eta^{\prime}\to\omega e^{+}e^{-}$ yield, an
unbinned maximum likelihood fit to
$M(\pi^0\pi^+\pi^-e^+e^-)-M(\pi^0\pi^+\pi^-)$, shown in
Fig.~\ref{fit3}, is performed.  The signal component is modeled by the
MC simulated signal shape convoluted with a Gaussian function to
account for the difference in the mass resolution between data and MC
simulation. The shape of the dominant non-resonant background
$\eta^\prime\to\pi^+\pi^-\eta$ is derived from the MC simulation, and
its magnitude is fixed taking into account the decay branching
fraction from the PDG~\cite{PDG}.  The remaining background contributions
are described with a 2nd-order Chebychev polynomial. The fit shown in
Fig.~\ref{fit3} yields $66\pm11$ $\eta^\prime\to \omega e^+e^-$ events
with a statistical significance of 8$\sigma$. The statistical
significance is determined by the change of the log-likelihood value and
of the number of degrees of freedom in the fit with and without the
$\eta^{\prime}\to\omega e^{+}e^{-}$ signal included.

To determine the detection efficiency, we produce a signal MC sample
in which $\eta^\prime\rightarrow \omega e^{+}e^{-}$ is modeled as the
decay amplitude in Ref.~\cite{Phys.Rev.C61.i} based on the VMD
model. After subtracting the peaking background events and taking into
account the detection efficiency of $(5.45\pm0.03)\%$, the branching
fraction of $\eta^{\prime}\to\omega e^{+}e^{-}$ is determined to be
$(1.97\pm0.34)\times10^{-4}$.  This is summarized in Table~\ref{sum}.

\begin{table}[htpb]
\begin{center}
  \caption{Signal yields, detection efficiencies and the branching
    fractions of $\eta^\prime\rightarrow \omega \gamma$ and
    $\eta^\prime\rightarrow \omega e^+e^-$. The first errors are statistical,
    and the second are systematical.}\label{sum}
\begin{tabular}{l|c|c|c}
\hline
 Decay mode  &  Yield  & $\varepsilon$(\%)& Branching fraction\\
\hline
$\eta^\prime\to\omega\gamma$
&  $33187\pm351$  & $21.87$   & $(2.55\pm0.03\pm0.16)\times10^{-2}$\\
$\eta^\prime\to \omega e^+e^-$& $66\pm 11$  & $5.45$
 & $(1.97\pm0.34\pm0.17)\times10^{-4}$\\
\hline

\end{tabular}
\end{center}
\end{table}

\section {Systematic uncertainties}
In this analysis, the systematic uncertainties on the branching fraction
measurements mainly come from the following sources:

\emph{a. MDC Tracking efficiency~~}

The tracking efficiencies of pions and electrons have been
investigated using clean samples of $J/\psi\rightarrow\rho\pi$,
$\psi^{\prime}\to\pi^{+}\pi^{-}J/\psi$, and $J/\psi\to e^{+}
e^{-}$($\gamma_{FSR}$).
Following the method described in Ref~\cite{photonerror}, we determine
the difference in tracking efficiency between data and simulation as 1\%
for each charged pion and 1.2\% for each electron. Therefore, 2\% is taken as the systematic
error of the tracking efficiency for $\eta^\prime\to\omega\gamma$ with
two charged tracks, and 4.4\% for $\eta^\prime\to\omega e^+e^-$ with
four charged tracks.

\emph{b. PID efficiency~~}

For $\eta^{\prime}\to\omega e^{+} e^{-}$, PID is used when we obtain
$\chi^{2}_{4C+PID}$ of every combination for each
event. The decay $J/\psi\to\pi^+\pi^{-}\pi^{0}$, with $\pi^{0}\to\gamma e^{+}
e^{-}$ is used as a control sample to estimate the difference
between data and MC with and without applying $\chi^{2}_{PID}$ to
identify the particle type. The difference, $3.8\%$, is taken as the
systematic uncertainty from PID for the decay $\eta^{\prime}\to\omega
e^{+} e^{-}$.

\emph{c. Photon detection efficiency~~}

The photon detection efficiency has been studied in $J/\psi\to\rho\pi$
decays in Ref.~\cite{photonerror}. The difference between data and MC simulation
is determined to be 1\% per photon. Therefore, 4\% and 3\% are
taken as the systematic uncertainties, respectively, for the two analyzed
$\eta^\prime$ decays.

\begin {table}[htp]
\caption{Summary of systematic uncertainties (in \%) for the branching
  fraction measurements. }\label{totalerror_br}
\begin{small}
\begin {tabular}{c c c c } \hline
Sources                 &$\eta^{\prime}\to\omega e^{+}e^{-}$  &$\eta^{\prime}\to\omega \gamma$
&$\frac{\mathcal{B}(\eta^{\prime}\to\omega e^{+}
e^{-})}{\mathcal{B}(\eta^{\prime}\to\omega \gamma)}$ \\  \hline \hline
    MDC tracking          &$4.4$                        &$2.0$      &$2.4$\\
    Photon detection        &$3.0$                      &$4.0$       &$1.0$\\
    PID                     &$3.8$                      &$-$      &$3.8$\\
    Kinematic fit           &$1.8$                      &$0.5$      &$1.9$\\
    $\gamma$ conversion subtraction  &$1.0$                 &$-$    &$1.0$\\
    Background uncertainty  &$3.7$                      &$2.9$   &$4.7$ \\
    Form factor uncertainty  &$1.3$                      &$-$      &$1.3$\\
    $\pi^{0}$ mass window     &$1.4$                     &$1.4$    &$-$\\
    $J/\psi$ total number    &$0.8$                       &$0.8$    &$-$\\
    $\mathcal{B}$($J/\psi\to\gamma\eta^{\prime}$)  &$3.1$     &$3.1$   &$-$ \\
    $\mathcal{B}$($\omega\to\pi^{0}\pi^{+}\pi^{-}$) &$0.8$     &$0.8$   &$-$ \\  \hline
    Total                   &$8.7$                         &$6.4$   &$7.0$\\  \hline
\end {tabular}
\end{small}
\end {table}

\emph{d. Kinematic fit~~}

The angular and momentum resolutions for charged tracks are significantly better
in simulation than in data.  This results in a narrower $\chi^2_{4c} $
distribution in MC than in data and introduces a systematic bias in the
efficiency estimation associated with the $4C$ kinematic fit.
The difference can be reduced by
correcting the track helix parameters of the simulated tracks, as described
in detail in Ref.~\cite{4cfiterror}.  In this analysis, a clean sample
of $J/\psi\to\pi^{+}\pi^{-}\pi^0,\pi^0\to\gamma e^+e^-$ is selected to
study the difference of the helix parameters of pions and electrons
between data and MC simulation. The helix parameters of each charged
track are corrected so that $\chi^2_{4C}$ from MC simulation is in
better agreement with that of data.  With the same correction factors,
the kinematic fit is performed for the signal MC events and the
$\chi^2_{4C}$ is required to be less than 80. By comparing the numbers
of selected signal events with and without the correction, we determine
the change in detection efficiencies to be $0.5\%$ and $1.8\%$.  These
are taken as the systematic uncertainties for $\eta^\prime\rightarrow
\omega\gamma $ and $\eta^\prime\rightarrow \omega e^+e^-$, respectively.

\emph{e. $\gamma$ conversion event veto~~}

In the analysis of $\eta^\prime\to \omega e^+e^-$, the large
contamination of $\gamma$ conversion events from the decay
$\eta^{\prime}\to\omega\gamma$ is effectively removed by the
requirement of $R_{xy}<2$~cm. To estimate the uncertainty
associated with this requirement, we select a clean sample of
$J/\psi\to\pi^{+}\pi^{-}\pi^{0}$ with $\pi^0\to\gamma e^+e^-$.
The efficiency corrected signal yields with and without the
$R_{xy}$ criterion differ by $1.0\%$, which is taken as the systematic
uncertainty.

\emph{f. Background~~}

The non-peaking background uncertainties in each channel are estimated
by varying the fit range and changing the background shape in the fit,
and they are determined to be 2.9\% for $\eta^\prime\to\omega\gamma$.
To reduce the statistical uncertainty for $\eta^{\prime}\to\omega e^{+} e^{-}$,
we use the background shape from the inclusive MC sample,
and the maximum change of the branching fraction, $3.6\%$ is taken as
the uncertainty from the non-peaking background.  In order to evaluate the background
uncertainty from $\eta^{\prime}\to\eta\pi^{+}\pi^{-}$ in the analysis of
the $\eta^{\prime}\to\omega e^{+} e^{-}$ decay, to, we perform an
alternative fit by varying its contribution according to the
uncertainty from branching fractions of $J/\psi\to\gamma\eta^\prime$
and its cascade decays.  We also vary the selection efficiency of this
background channel as determined by the MC sample, and find that the total difference
in the signal yield is about 0.3\%, which can be ignored. In addition, the
change in the number of peaking background events from
$\eta^\prime\to\omega\gamma$ due to a difference of the $\gamma$
conversion ratio between MC and data leads to an uncertainty of 1.0\%
on the branching fraction of $\eta^\prime\to \omega e^+e^-$.
The total background uncertainties from these sources are listed in
Table.~\ref{totalerror_br}.

\emph{g. Form factor~~}

The nominal signal MC model is based on the amplitude in Ref.~\cite{Phys.Rev.C61.i}
To
evaluate the uncertainty due to the choice of the form factors in the
determination of the detection efficiency, we also generate MC
samples with other form factors in Ref.~\cite{Phys.Rev.C61.i}, \textit{e.g.},
the monopole and dipole parameterizations. The maximum change of the
detection efficiency, $1.3\%$, is regarded as the systematic
uncertainty from this source.

\emph{h. $\pi^{0}$ mass window requirement}

The uncertainty from the $\pi^{0}$ mass window requirement due to the
difference in the mass resolution between data and simulation is
estimated by comparing the difference in efficiency of $\pi^{0}$ invariant mass
window requirement between data and signal MC
simulation. It is determined to be $1.4\%$ for the
$\eta^{\prime}\to\omega \gamma$ mode.
Since the $\pi^0$ kinematics in the $\eta^{\prime}\to\omega e^{+}e^{-}$ decay
is similar to the $\eta^{\prime}\to\omega \gamma$ mode, the same value is
taken as the uncertainty from this source for both decay modes.

The contributions of systematic uncertainties studied above and the
uncertainties from the branching fractions
($J/\psi\to\gamma\eta^\prime$ and $\omega\to\pi^+\pi^-\pi^0$) and the
number of $J/\psi$ events are summarized in Table~\ref{totalerror_br},
where the total systematic uncertainty is obtained by adding the
individual contributions in quadrature, assuming all sources to be
independent.

\section{Results}
The signal yields and detection efficiencies used to calculate the
branching fractions and the corresponding results are listed in
Table.~\ref{sum}. Using the PDG world averages of
$\mathcal{B}(J/\psi\to\gamma\eta^\prime)$ and
$\mathcal{B}(\omega\to\pi^{0}\pi^{+}\pi^{-})$~\cite{PDG}, the
branching fractions of $\eta^\prime\to\omega \gamma $ and
$\eta^\prime\to \omega e^+e^-$ are determined to be
$\mathcal{B}(\eta^{\prime}\to \omega \gamma)=
(2.55\pm0.03$(stat)$\pm0.16$(syst)$)\times10^{-2}$ and
$\mathcal{B}(\eta^{\prime}\to \omega
e^{+}e^{-})=(1.97\pm0.34$(stat)$\pm0.17$(syst)$)\times10^{-4}$,
respectively. The ratio $\frac{\mathcal{B}(\eta^{\prime}\to \omega
e^{+} e^{-})} {\mathcal{B}(\eta^{\prime}\to \omega \gamma)}$ is then
determined to be $(7.71\pm1.34$(stat)$\pm0.54$(syst)$)\times10^{-3}$,
where several systematic uncertainties cancel, \textit{e.g.}, the uncertainties
associated with the charged pions (MDC tracking), photon detection efficiency,
branching fractions of $J/\psi\to\gamma\eta^\prime$ and
$\omega\to\pi^+\pi^-\pi^0$ and the $\pi^{0}$ mass window requirement.

\section{Summary}
With a sample of $1.31$ billion $J/\psi$ events collected with the
BESIII detector, we have analyzed the decays $\eta^{\prime}\to\omega\gamma$ and
$\eta^{\prime}\to\omega e^{+} e^{-}$ via
$J/\psi\to\gamma\eta^\prime$. For the first time, the decay of
$\eta^{\prime}\to\omega e^{+} e^{-}$ is observed with a statistical significance
of 8$\sigma$, and its branching fraction is measured to be
$\mathcal{B}(\eta^{\prime}\to\omega
e^{+}e^{-})=(1.97\pm0.34$(stat)$\pm0.17$(syst)$)\times10^{-4}$, which
is consistent with theoretical prediction,
$2.0{\times}10^{-4}$~\cite{Phys.Rev.C61.i}. The branching fraction of
$\eta^\prime\to \omega\gamma$ is determined to be
$\mathcal{B}(\eta^{\prime}\to \omega \gamma)=
(2.55\pm0.03$(stat)$\pm0.16$(syst)$)\times10^{-2}$, which is in good
agreement with the world average value in Ref.~\cite{PDG} and the most
precise measurement to date. In addition, the ratio
$\frac{\mathcal{B}(\eta^{\prime}\to \omega e^{+} e^{-})}
{\mathcal{B}(\eta^{\prime}\to \omega \gamma)}$ is determined to be
$(7.71\pm1.34$(stat)$\pm0.54$(syst)$)\times10^{-3}$.

\section*{Acknowledgements}
The BESIII collaboration thanks the staff of BEPCII and the IHEP computing center for their strong support. This work is supported in part by National Key Basic Research Program of China under Contract No. 2015CB856700; National Natural Science Foundation of China (NSFC) under Contracts Nos. 11125525, 11235011, 11322544, 11335008, 11425524, 11175189; Youth Science Foundation of China under constract No. Y5118T005C; the Chinese Academy of Sciences (CAS) Large-Scale Scientific Facility Program; the CAS Center for Excellence in Particle Physics (CCEPP); the Collaborative Innovation Center for Particles and Interactions (CICPI); Joint Large-Scale Scientific Facility Funds of the NSFC and CAS under Contracts Nos. 11179007, U1232201, U1332201; CAS under Contracts Nos. KJCX2-YW-N29, KJCX2-YW-N45; 100 Talents Program of CAS; National 1000 Talents Program of China; INPAC and Shanghai Key Laboratory for Particle Physics and Cosmology; German Research Foundation DFG under Contract No. Collaborative Research Center CRC-1044; Istituto Nazionale di Fisica Nucleare, Italy; Ministry of Development of Turkey under Contract No. DPT2006K-120470; Russian Foundation for Basic Research under Contract No. 14-07-91152; The Swedish Resarch Council; U. S. Department of Energy under Contracts Nos. DE-FG02-04ER41291, DE-FG02-05ER41374, DE-FG02-94ER40823, DESC0010118; U.S. National Science Foundation; University of Groningen (RuG) and the Helmholtzzentrum fuer Schwerionenforschung GmbH (GSI), Darmstadt; WCU Program of National Research Foundation of Korea under Contract No. R32-2008-000-10155-0.

\end{document}

%% file: authors_jul2015.tex
\author{
  \begin{small}
    \begin{center}
      M.~Ablikim$^{1}$, M.~N.~Achasov$^{9,f}$, X.~C.~Ai$^{1}$,
      O.~Albayrak$^{5}$, M.~Albrecht$^{4}$, D.~J.~Ambrose$^{44}$,
      A.~Amoroso$^{49A,49C}$, F.~F.~An$^{1}$, Q.~An$^{46,a}$,
      J.~Z.~Bai$^{1}$
% ...
      \\
      \vspace{0.2cm}
      (BESIII Collaboration)\\
      \vspace{0.2cm} {\it
        $^{1}$ Institute of High Energy Physics, Beijing 100049, People's Republic of China\\
% ...
$^{4}$ Bochum Ruhr-University, D-44780 Bochum, Germany\\
        $^{5}$ Carnegie Mellon University, Pittsburgh, Pennsylvania 15213, USA\\
% ...
$^{9}$ G.I. Budker Institute of Nuclear Physics SB RAS (BINP), Novosibirsk 630090, Russia\\
% ...
$^{44}$ University of Rochester, Rochester, New York 14627, USA\\
% ...
$^{46}$ University of Science and Technology of China, Hefei 230026, People's Republic of China\\
% ...
$^{49}$ (A)University of Turin, I-10125, Turin, Italy; (B)University of Eastern Piedmont, I-15121, Alessandria, Italy; (C)INFN, I-10125, Turin, Italy\\
% ...
$^{a}$ Also at State Key Laboratory of Particle Detection and Electronics, Beijing 100049, Hefei 230026, People's Republic of China\\
% ...
$^{f}$ Also at the Novosibirsk State University, Novosibirsk, 630090, Russia\\
% ...
      }
    \end{center}
  \end{small}
}

\affiliation{}